\documentclass[%
 reprint,
 amsmath,amssymb,
 aps,
]{revtex4-2}

\usepackage{graphicx}
\usepackage{dcolumn}
\usepackage{bm}
\usepackage{mathrsfs}
\usepackage{float}      
\usepackage[table]{xcolor}      
\usepackage{subcaption} 
  
\usepackage{multirow}
\usepackage{makecell}
 \usepackage{amsmath}
\usepackage{graphicx}
\usepackage{url}
\usepackage{fancyhdr}
\usepackage{footmisc}
\usepackage{algorithmic}
\usepackage{listings}
\usepackage{fancyvrb}
\usepackage{graphicx}
\usepackage{hyperref}
 
\begin{document}

\graphicspath{ {./} }
\preprint{APS/123-QED}

\title{Exact diagonalization for spin-1/2 spin ice pyrochlores}

\author{C.~Wei}
\email{cw7734@mun.ca}
\author{S.~H.~Curnoe}%
 \email{curnoe@mun.ca}
\affiliation{%
 Department of Physics and Physical Oceanography, Memorial University of Newfoundland,
St. John's, Newfoundland \& Labrador, Canada A1B 3X7
}%

\date{\today}

\begin{abstract}
We find exact solutions to the Hamiltonian of a 16-site spin-1/2 pyrochlore crystal with nearest neighbour exchange interactions.  The methods of group theory (symmetry) are used to completely block-diagonalize the Hamiltonian, yielding precise details about symmetry of the eigenstates, in particular those components which are {\em spin ice} states, in order to evaluate the spin ice density at finite temperature. At low enough temperatures, a `perturbed' spin ice phase is clearly outlined within the four parameter space of the general model of exchange interactions. The quantum spin ice phase is expected to exist outside these boundaries.
\end{abstract}

                              
\maketitle

\section{Introduction}\label{sec:Intro}
The ideal spin ice was proposed by Philip Anderson in 1956\cite{anderson_physrev_1956} in a
study of finite entropy, short-range ordered states in magnetic spinels. In pyrochlores and spinels (which share the same crystallographic symmetry group,
$Fd{\bar 3}m$),  the ``spin ice rule" is that the four Ising-like spins on a tetrahedron must be oriented so that two of them point toward the centre of each tetrahedron, and two point away, the so-called ``two-in-two-out" configuration.\cite{anderson_physrev_1956}  Given that the tetrahedra share spins on each vertex, the number of pure spin ice states (those states for which the spin ice rule is satisfied on {\em every} tetrahedron) is constrained by the geometry of the system and is approximately  $1.5^{N/2}$ \cite{pauling1935}.  The spin ice rule is precisely analogous to the ice rule for water ice, which governs the positions of the hydrogen atoms, and which results in a highly degenerate ground state in water ice, producing a residual entropy at zero  
temperature, as shown by Pauling in 1935 \cite{pauling1935}; spin ice is expected to have the same property \cite{anderson_physrev_1956}.  This classical spin ice state occurs in several pyrochlores, including Ho$_2$Ti$_2$O$_7$ and Dy$_2$Ti$_2$O$_7$ \cite{harris_1997_PhysRevLett.79.2554, Ramirez_nature_1999}. The elementary excitations of spin ice are magnetic monopoles, which occur when the spin ice rule is violated and behave in accordance with Maxwell's equations \cite{Castelnovo_nature_2008, Jaubert_Holdsworth_peter_nat_phys_2009}.

In contrast to the classical spin ice materials, a quantum spin ice occurs when there are quantum fluctuations of the degenerate states even at zero temperature, which may lead to the formation of a quantum spin liquid \cite{hermele2004_PhysRevB.69.064404,shannon_2012_PhysRevLett.108.067204, benton_2012_PhysRevB.86.075154, Balents_nature_2010}.  Examples of candidate quantum spin ice pyrochlores include Tb$_2$Ti$_2$O$_7$\cite{Gardne_physrevb_2001, curnoe_2013_PhysRevB.88.014429}, Yb$_2$Ti$_2$O$_7$ \cite{Ross_physrevX_2011,chang_2021_nature_commun, pan_nature_2014}, Ce$_2$Zr$_2$O$_7$ \cite{gaudet_2019_PhysRevLett.122.187201}, 
and several Pr oxides 
\cite{zhou_2008_PhysRevLett.101.227204, robert_2015_PhysRevB.92.064425, sibille_2016_PhysRevB.94.024436, petit_2016_PhysRevB.94.165153}.



In this article we investigate a  quantum spin Hamiltonian for the pyrochlore lattice using exact diagonalization. This numerical method can be used to evaluate the entire spectrum, as well as finite temperature spin-correlations, for small systems. 
This method has been applied to spin-1/2  pyrochlore magnets with anti-ferromagnetic interactions \cite{Chandra_physrevb_2018} in order to investigate spin correlations on the ``breathing" pyrochlore lattice, which occurs when the tetrahedra of two different orientations have two different sets of coupling constants. In this work, we study the phase space of the most general model of exchange interactions on the pyrochlore lattice in the vicinity of the (ferromagnetic) spin ice phase.
The general form of the Hamiltonian has four exchange parameters and the methods of group theory are used to block-diagonalise the Hamiltonian in order to reduce the computational effort required to investigate this 4-parameter space.  One of our aims is to determine the position of candidate quantum spin ice materials within this space.



\section{Perturbed Spin Ice }\label{sec:Model}

\subsection{The Hamiltonian}
In pyrochlore magnets the spins occupy the 16d Wyckoff position of 
the crystallographic space group $Fd\bar{3}m$, which are the vertices of a network of corner-sharing tetrahedra.  The site symmetry is $D_{3d}$ and there are four inequivalent sites
(which we number $s=1,2,3,4$) which are distinguished by the direction of their 3-fold symmetry axes.
The tetrahedra alternate between two orientations, often called `$A$' and `$B$'.  Each spin is shared between an $A$
and a $B$ tetrahedron, and there are four tetrahedra of each type within a face-centred cubic cell.  Spin states can be written as 
kets of the form 
$|\pm\pm\pm \ldots\rangle$ where the quantization axis of each spin is assumed to lie along
the direction of its 3-fold symmetry axis, which points toward or away from the centres of the tetrahedra which share the spin. For example, the `all-in-all-out' state $|+++ \ldots \rangle$ has all spins pointing
{\em out of} one set of tetrahedra (say the `$A$` set) and {\em into} the other set. Two spins located on the same tetrahedron (of either orientation) are 
nearest neighbours. 

The $D_{3d}$ site symmetry lifts the $2J+1$-fold degeneracy of spin $J$ magnetic ions into singlets and doublets, which can be either Kramers (1/2-integral $J$) or non-Kramers (integral $J$). All cases where there is a well-separated doubly degenerate ground state can be treated as effective spin-1/2 systems \cite{curnoe_2013_PhysRevB.88.014429, curnoe_condensedmatter_2018}; thus we consider a general nearest-neighbor exchange interaction for spin-1/2 spins,
\begin{equation}
    H_{ex}=\sum_{\langle i,j\rangle}{\cal J}_{i,j}^{\mu \nu}S^\mu_i S^\nu_j,
\end{equation}
where the sum over $\langle i,j\rangle$ runs over pairs of nearest-neighbour spins
and $\vec{S}_i=(S^x_i, S^y_i,S^z_i)$ is the spin operator for the $i$th site. ${\cal J}_{i,j}^{\mu \nu}$ are exchange constants which are constrained by the space group symmetry of the crystal. In pyrochlore magnets
there are only four independent parameters. 
It is convenient to express the Hamiltonian as
\begin{equation}
H = {\cal J}_{1}X_1+{\cal J}_{2}X_2+{\cal J}_{3}X_3+{\cal J}_{4}X_4,
\label{eq:hamil}
\end{equation}
where ${\cal J}_{a}$ are the exchange constants and
\begin{align*}
&X_1=-\frac{1}{3}\sum_{\langle i,j \rangle}S_{iz}S_{jz}\\
&X_2=-\frac{\sqrt{2}}{3}\sum_{\langle i,j\rangle}[\Lambda_{s_is_j}(S_{iz}S_{j+}+S_{jz}S_{i+})+{\rm h.c.}]\\
&X_3=\frac{1}{3}\sum_{\langle i,j\rangle}[\Lambda_{s_is_j}^*S_{i+}S_{j+}+ {\rm h.c.}]\\
&X_4=-\frac{1}{6}\sum_{\langle i,j \rangle}(S_{i+}S_{j-}+ {\rm h.c.}).
\end{align*}
In these expressions, the spin operators $\vec{S}_i$ are given in terms of a set of local axes, such that the local $z$ axes are the spin quantization axes described above (see Refs.\ \cite{Curnoe2007_PhysRevB_2007, Curnoe_physrevb_2008} for
more details). $\Lambda_{ss'}$ are phases which depend on the site numbers: $\Lambda_{12}=\Lambda_{34}=1$ and $\Lambda_{13}=\Lambda_{24}=\Lambda_{14}^{*}=\Lambda_{23}^{*}=\varepsilon\equiv \exp(\frac{2\pi i}{3})$, $S_{\pm}=S_x\pm i S_y$.
Note that when ${\cal J}_{1}={\cal J}_{2}={\cal J}_{3}={\cal J}_{4} \equiv {\cal J}$ the exchange interaction is simply the isotropic (Heisenberg) exchange, $H_{\rm iso} = {\cal J}\sum\limits_{\langle i,j \rangle}\Vec{J_{i}}\cdot \Vec{J_{j}}$ \cite{Curnoe_physrevb_2008}.
In the special case when 
${\cal J}_2 = {\cal J}_3 = {\cal J}_4 = 0$ all states of the form $|\pm \pm \pm \ldots \rangle$ ({\em i.e.}  the basis kets)
are eigenstates of $H$. If ${\cal J}_1 >0$ then the ground state will be the doubly 
degenerate ``all-in-all-out" states
$|+++ \ldots\rangle$ and $|--- \ldots \rangle$, 
otherwise the ground state is the set of 
highly degenerate ``two-in-two-out" spin ice states. 
We examine perturbations around the spin ice state by finding exact numerical solutions for $H$ for ${\cal J}_1 >0$ and the
other exchange constants small ($|{\cal J}_{2,3,4}| \ll {\cal J}_1$), with the aim of determining the range of existence of spin ice  as the system evolves into a quantum spin ice. 

\subsection{Finite Temperature}
To study the system at finite temperature we consider the density function, 
\begin{eqnarray}
    \rho & = & \frac{1}{Z} \exp -H/T \nonumber \\
    & = & 
    \frac{1}{Z}\sum_{i} \text{exp} (-E_i/T)|\psi_i\rangle\langle\psi_i| 
\end{eqnarray}
where $|\psi_i\rangle$ are the eigenstates of $H$ with eigenvalues $E_i$ and \begin{equation}
    Z=\sum_{i} \text{exp}(-E_i/T)
\end{equation}
is the partition function.

Each basis state $|\pm\pm\pm \ldots\rangle$ is a unique spin configuration on the 8 tetrahedra ($A$-type and $B$-type) of our 16-site system.  For each state, the number of all-in-all-out tetrahedra is $N_{AIAO}$, the number of two-in-two-out (ice rule) tetrahedra is $N_{2I2O}$, and the number of three-in-one-out or one-in-three-out tetrahedra is $N_{31/13}$, where
$N_{AIAO} + N_{2I2O} + N_{31/13} =8$.  A pure spin ice state satisfies the spin ice rule on all tetrahedra and will have $N_{2I2O} = 8$. 
Thus the total density of each configuration is
\begin{equation}
n_{config}  =  \sum_j \left(\frac{N_{config,j}}{8} \right) \langle u_j |\rho|u_j\rangle
\label{spin_ice}
\end{equation}
where $|u_j\rangle$ are the basis states. 


\subsection{Method}

Generally, the Hamiltonian can be block-diagonalized into sectors classified by the irreducible representations (IR's) of the underlying space group. These will be labelled by a $k$-vector to which will belong several different IR's. 
In a finite system with periodic boundary conditions only a small number of $k$-vectors
will occur.  
In order to fully exploit the 
symmetry of the system, the shape of the finite system should be chosen so that the point group symmetry is preserved along with a set of translations. Periodic boundary conditions are assumed.
%
%
The smallest unit that preserves the point group symmetry is a single tetrahedron with four spins; here there are no translations and the only $k$-vector is $k = (0,0,0)$, the $\Gamma$ point of the Brillouin zone. The next smallest unit has 16 spins arranged on four tetrahedra inside a cubic cell; here the IR's belong to  the $\Gamma$-point or the $X$-point, $k=\frac{2\pi}{a}(1,0,0)$.  

The symmetry group of the sixteen-site system contains 192 elements 
with 14 IR's (10 belonging to the 
$\Gamma$-point and 4 to the $X$-point), of which all but four are degenerate. Thus the 
$2^{16}\times 2^{16}$ Hamiltonian matrix can be block-diagonalized 
by an appropriate unitary transformation 
(see  Appendix A). Among the $2^{16}$ basis states, only 90 are pure spin ice states
which occur in 12 of the 14 IR's.


\section{Results}\label{sec:Discuss}

\begin{figure}[H]
\centering
\subfloat[${\cal J}_{1} = -1$,  ${\cal J}_{4}=0$]{
  \includegraphics[width=40mm]{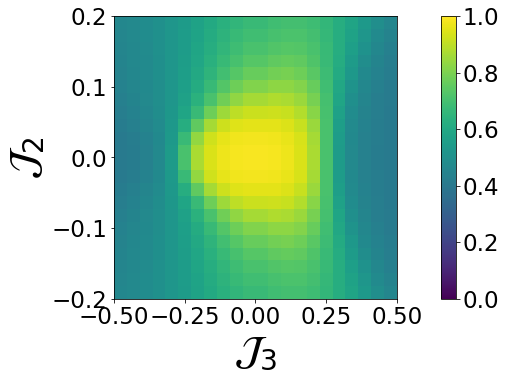}
}
\subfloat[${\cal J}_{1} = -1$, ${\cal J}_{3}=0$]{
  \includegraphics[width=40mm]{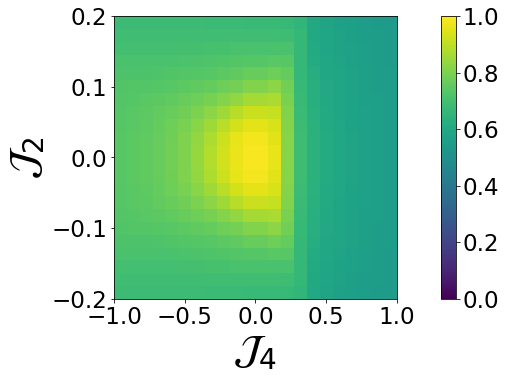}
}
\hspace{0mm}
\subfloat[${\cal J}_{1} = -1$, ${\cal J}_{2} = 0$]{
  \includegraphics[width=40mm]{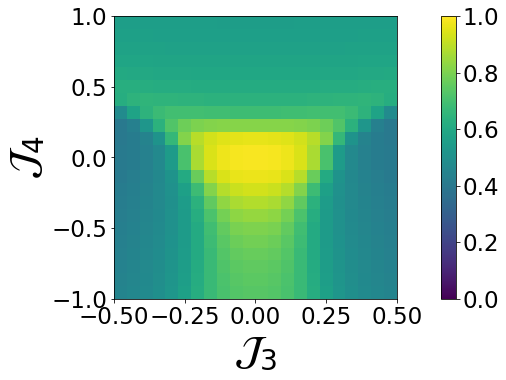}
}
\subfloat[${\cal J}_{1} = -1$]{
  \includegraphics[width=40mm]{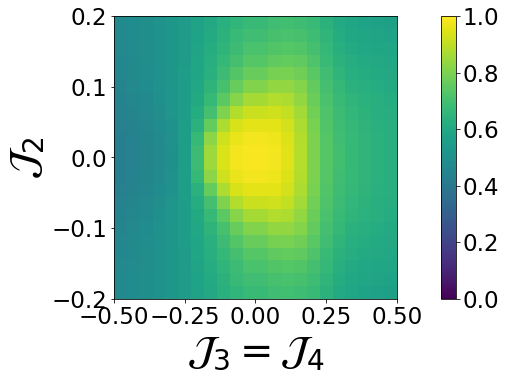}
}

\caption{The two-in-two-out density $n_{2I2O}$ at  $T=0.05$. 
\label{fig:16-1}}
\end{figure}

\begin{figure}[H]
\centering
\subfloat[${\cal J}_{1} = -1$, ${\cal J}_{4} =0$]{
  \includegraphics[width=40mm]{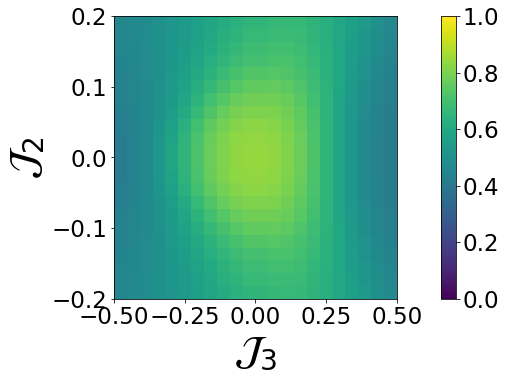}
}
\subfloat[${\cal J}_{1} = -1$, ${\cal J}_{3}=0$]{
  \includegraphics[width=40mm]{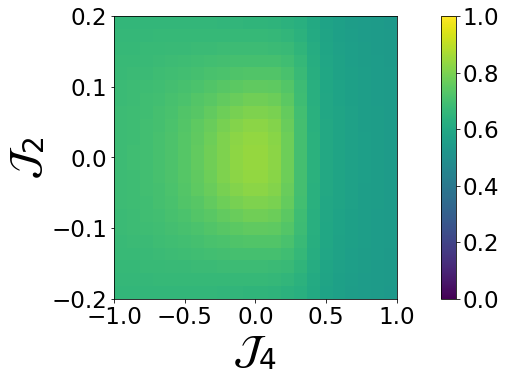}
}
\hspace{0mm}
\subfloat[${\cal J}_{1} = -1$, ${\cal J}_{2}=0$]{
  \includegraphics[width=40mm]{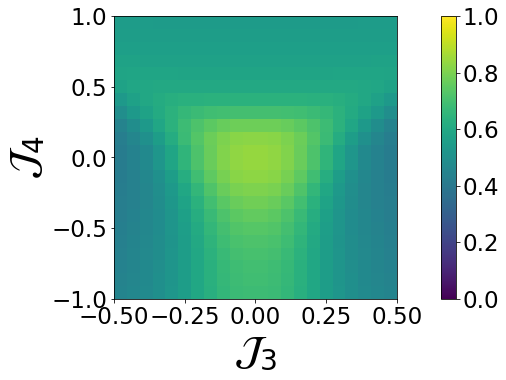}
}
\subfloat[${\cal J}_{1}$ = -1]{
  \includegraphics[width=40mm]{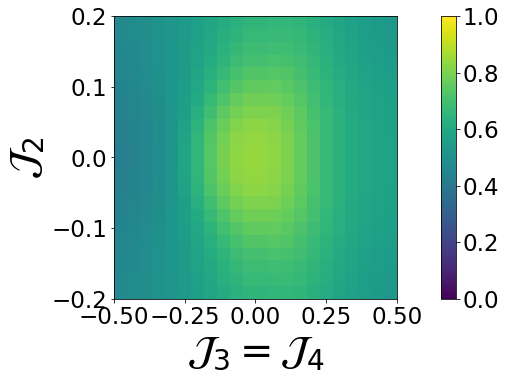}
}

\caption{The two-in-two-out density $n_{2I2O}$ at $T=0.1$. 
\label{fig:16-2}}
\end{figure}

\begin{figure}[H]
\centering
\subfloat[${\cal J}_{1} = -1$,  ${\cal J}_{4}=0$]{
  \includegraphics[width=40mm]{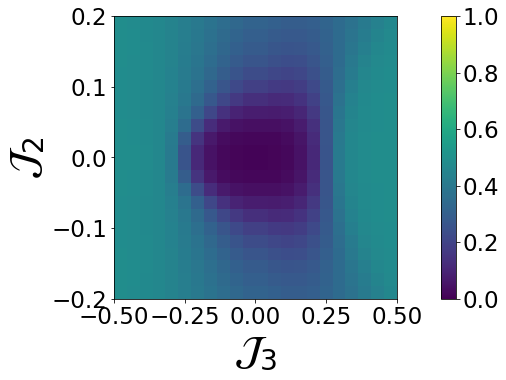}
}
\subfloat[${\cal J}_{1} = -1$, ${\cal J}_{3}=0$]{
  \includegraphics[width=40mm]{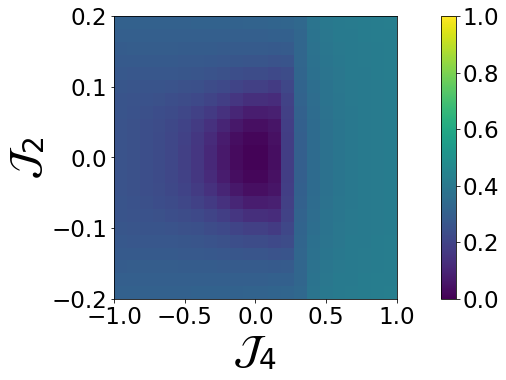}
}
\hspace{0mm}
\subfloat[${\cal J}_{1} = -1$, ${\cal J}_{2} = 0$]{
  \includegraphics[width=40mm]{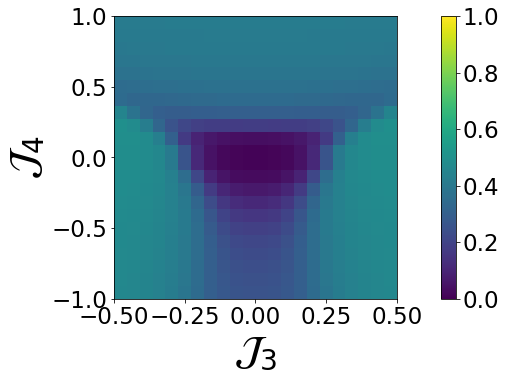}
}
\subfloat[${\cal J}_{1} = -1$]{
  \includegraphics[width=40mm]{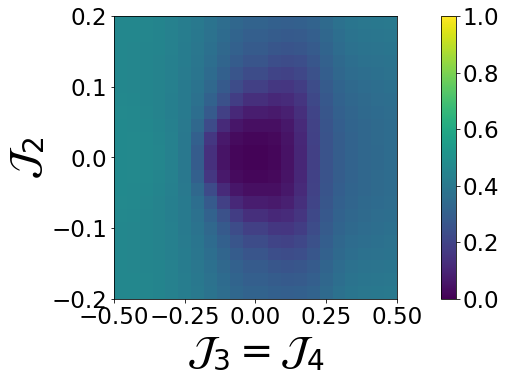}
}

\caption{The three-in-one-out/one-in-three-out density $n_{31/13}$ at  $T=0.05$. 
\label{fig:16-3}}
\end{figure}

\begin{figure}[H]
\centering
\subfloat[${\cal J}_{1} = -1$,  ${\cal J}_{4}=0$]{
  \includegraphics[width=40mm]{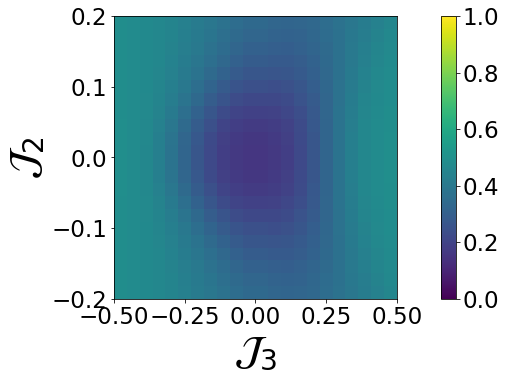}
}
\subfloat[${\cal J}_{1} = -1$, ${\cal J}_{3}=0$]{
  \includegraphics[width=40mm]{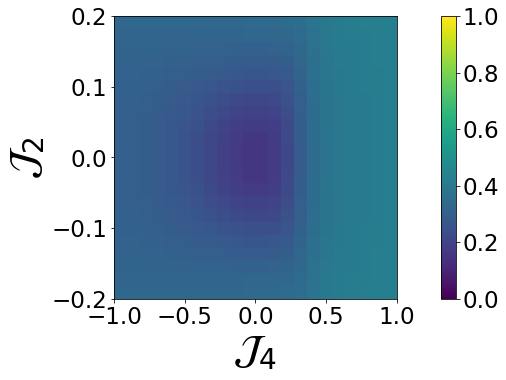}
}
\hspace{0mm}
\subfloat[${\cal J}_{1} = -1$, ${\cal J}_{2} = 0$]{
  \includegraphics[width=40mm]{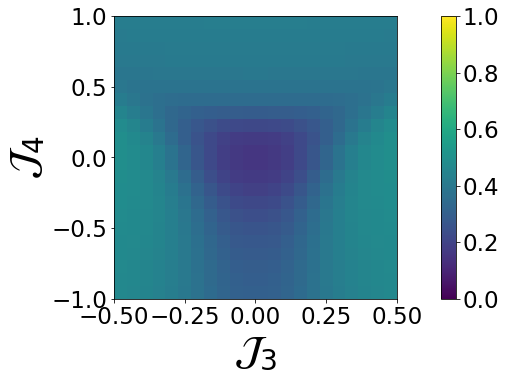}
}
\subfloat[${\cal J}_{1} = -1$]{
  \includegraphics[width=40mm]{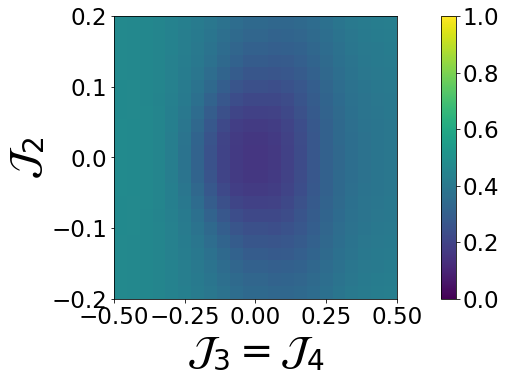}
}

\caption{The three-in-one-out/one-in-three-out density $n_{31/13}$ at  $T=0.1$. 
\label{fig:16-4}}
\end{figure}

\begin{figure}[H]
\centering
\subfloat[${\cal J}_{1} = -1$,  ${\cal J}_{4}=0$]{
  \includegraphics[width=40mm]{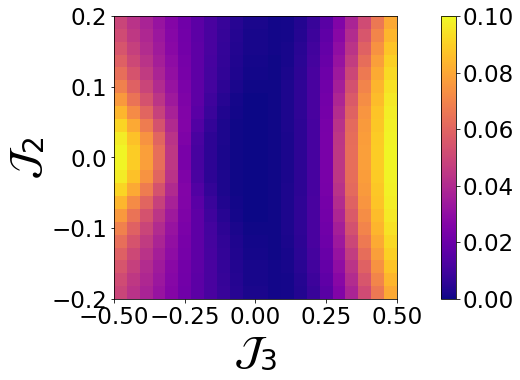}
}
\subfloat[${\cal J}_{1} = -1$, ${\cal J}_{3}=0$]{
  \includegraphics[width=40mm]{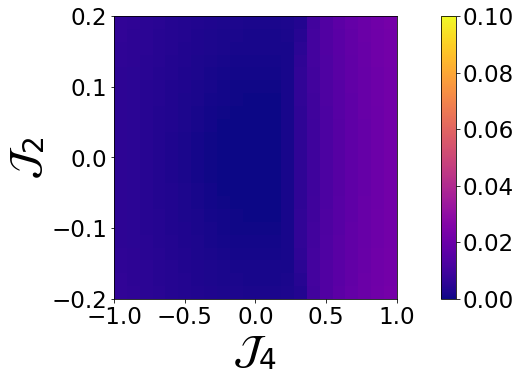}
}
\hspace{0mm}
\subfloat[${\cal J}_{1} = -1$, ${\cal J}_{2} = 0$]{
  \includegraphics[width=40mm]{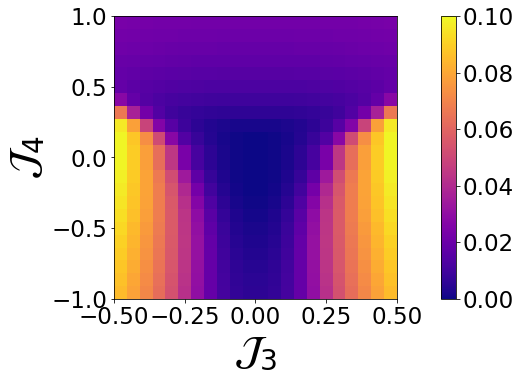}
}
\subfloat[${\cal J}_{1} = -1$]{
  \includegraphics[width=40mm]{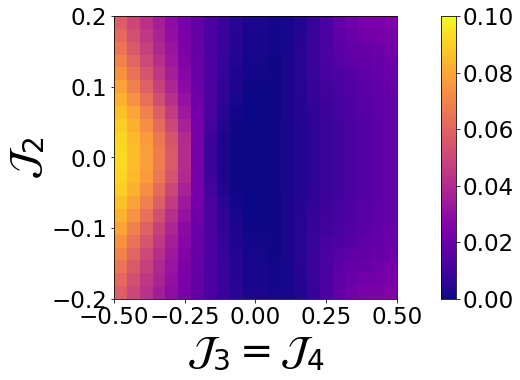}
}

\caption{The all-in-all-out density $n_{AIAO}$ at  $T=0.05$. 
\label{fig:16-5}}
\end{figure}

\begin{figure}[H]
\centering
\subfloat[${\cal J}_{1} = -1$,  ${\cal J}_{4}=0$]{
  \includegraphics[width=40mm]{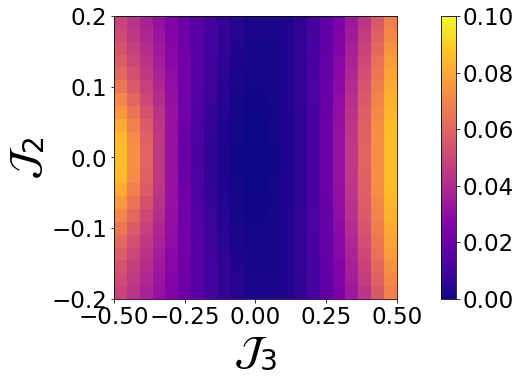}
}
\subfloat[${\cal J}_{1} = -1$, ${\cal J}_{3}=0$]{
  \includegraphics[width=40mm]{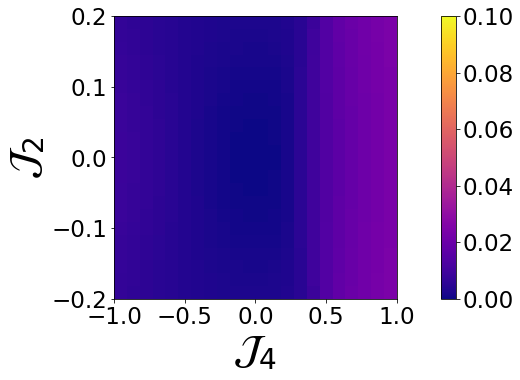}
}
\hspace{0mm}
\subfloat[${\cal J}_{1} = -1$, ${\cal J}_{2} = 0$]{
  \includegraphics[width=40mm]{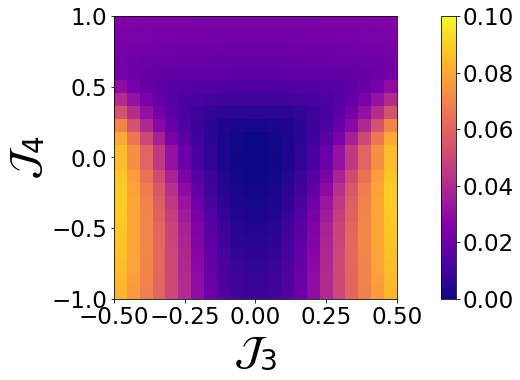}
}
\subfloat[${\cal J}_{1} = -1$]{
  \includegraphics[width=40mm]{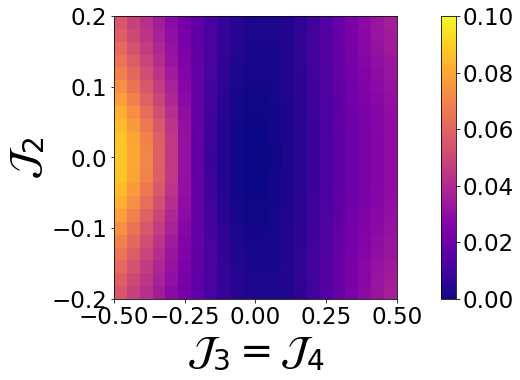}
}

\caption{The all-in-all-out density $n_{AIAO}$ at  $T=0.1$. 
\label{fig:16-6}}
\end{figure}

We solved the block-diagonalized $H$ numerically for 
${\cal J}_1 = -1$ and computed the densities $n_{AIAO}$, $n_{2I2O}$ and $n_{31/13}$ (Eq.\ \ref{spin_ice}) for select planes within the parameter space of
${\cal J}_2$, ${\cal J}_3$ and ${\cal J}_4$. The results are shown in Figs.~ \ref{fig:16-1}-\ref{fig:16-6} at two different temperatures. Figs.\ \ref{fig:16-1} and \ref{fig:16-2} show the two-in-two-out density $n_{2I2O}$. A well-defined, finite region in the parameter space in which the ground state is predominately spin ice is evident. Outside this region, the number of tetrahedra satisfying the spin ice rule is between roughly 50\% and 80\%.


%

Figs.\ \ref{fig:16-3} and \ref{fig:16-4} show the three-in-one-out/one-in-three-out density
$n_{31/13}$ using the same parameter ranges, temperatures and overall scale as the two-in-two-out results above. Figs.\ \ref{fig:16-5} and \ref{fig:16-6} show the all-in-all-out density
$n_{AIAO}$ at the same parameter ranges and temperatures, but a different colour scheme has been used because the maximum density is only 10\%, and is negligeable for much of the parameter space. Hence the plots of $n_{31/13}$ mirror those of $n_{2I2O}$, with a well-defined region centred at the origin where the density vanishes, and a maximum value of approximately 50\% outside this region.

The origin point of the figures, corresponding to ${\cal J}_2 = {\cal J}_3 = {\cal J}_4=0$ and ${\cal J}_1 = -1$ can be solved analytically, yielding a 90-fold degenerate pure spin ice 
ground state with energy of $-4/3$ and a 2784-fold degenerate first excited state
with energy $-1$. The spacing between energy levels is 1/3, up to the highest energy all-in-all-out state with energy 2.
Generally, the energy of the pure spin ice state is $-N/12$, where $N$ is the number of sites and $N/2$ is the total number of tetrahedra, while the energy of the highest level is $N/4$ and the minimum level spacing is $1/3$. 

The notable feature of our results is the extent of the pure spin ice phase for well-defined, finite ranges of the constants ${\cal J}_2,$  ${\cal J}_3$ and ${\cal J}_4$.
In Appendix B we show the results for a smaller system - a single tetrahedron. We find that in the smaller system the range of the spin ice ground state as seen in the spin ice density plots is similar to the larger system for the parameters ${\cal J}_2$ and ${\cal J}_3$, but finite size effects for the single-tetrahedron are pronounced when ${\cal J}_4$ is non-zero. Finite size effects arise when the number of basis states is small compared to the number of irreducible representations, which occurs in the four-site single-tetrahedron problem, but not in the larger sixteen-site problem.

In praseodymium pyrochlores the constant ${\cal J}_2$ is zero because the term $X_2$ in $H$ is not invariant under time reversal for the non-Kramers ground state doublet of the Pr ions \cite{onoda_2011_PhysRevB.83.094411}. The phase diagram for this model was studied using mean field theory in Ref.\ \cite{petit_2016_PhysRevB.94.165153}, with the boundaries of the spin ice phase for similar to what we have shown in Figs.\ 1(c) and 2(c); the roughly triangular shape of the central region is similar, as are the other phase boundaries appearing outside this region.  (The relations between the coupling constants used in Ref.\ \cite{petit_2016_PhysRevB.94.165153} and those used in Eq.\ \ref{eq:hamil} are given in Appendix C). Outside this region, Ref.\ \cite{petit_2016_PhysRevB.94.165153} identifies  `quantum' phases associated with spins arranged perpendicular to their local $z$-axes, which are not incompatible with the superposition of three-in-one-out/one-in-three-out and two-in-two-out states evident in our results.

We can determine the location in our phase diagrams of quantum spin ice candidates using estimates of coupling constants.  For Yb$_2$Ti$_2$O$_7$, the exchange constants were determined by fits to spin wave spectra, ${\cal J}_1 = -5.9$ K, and 
${\cal J}_2/|{\cal J}_1| = 0.17$, 
${\cal J}_3/|{\cal J}_1| = 0.29$ and
${\cal J}_4/|{\cal J}_1| = 0.05$ \cite{Ross_physrevX_2011}, corresponding to a location that is just outside the pure spin ice region but where the spin ice component is still large (about 70\%), consistent with a `quantum' spin ice state. 

The terbium ions in Tb$_2$Ti$_2$O$_7$ are another example of non-Kramers doublets, but mixing with excited states (within the $2J+1$ multiplet) and a symmetry-preserving map allows for all four exchange terms to be present in the Hamiltonian.
Assuming a perturbative renormalization of the coupling constants, estimates of them obtained from fits to diffuse neutron scattering measurements are:
${\cal J}_1 = -5.1$ K, ${\cal J}_2/|{\cal J}_1| = .04$,
${\cal J}_3/|{\cal J}_1| = .02$ and ${\cal J}_4/|{\cal J}_1| = .06$ \cite{curnoe_2013_PhysRevB.88.014429}, which is near the centre of the pure spin ice region, suggesting that this material is a perturbed spin ice. 


\section{Summary}\label{sec:Discuss}

Useful physical insight, in addition to computational advantages, can be gained by exploiting the high symmetry underlying pyrochlore magnets. Here we have combined this approach with numerical methods to determine precisely how the spin ice contribution to the total density function varies according to the coupling constants of the most general model for nearest-neighbour spin-spin interactions. A well-defined region centered at the origin of this parameter space is the location of a phase which can be thought of as a perturbed spin ice - a phase which is largely a pure, classical spin ice with small fluctuations due to the transverse terms $X_2$, $X_3$ and $X_4$ in the nearest neighbour exchange Hamiltonian. Beyond this region, where transverse fluctuations are larger,  boundaries between other phases are also evident, suggestive of competing quantum phases.


\begin{acknowledgments}
We thank Oliver Stueker for assistance with
using the resources at Compute Canada and
Kyle Hall for helpful discussions about coding. 
	\end{acknowledgments}

	\appendix


\section{Block Diagonalization}

The pyrochlore crystal structure has space group symmetry $Fd\Bar{3}m$ (No.\ 227), with octahedral point group symmetry $O_h$ and  fcc lattice translations $n_1\vec{t}_1+n_2\vec{t}_2+n_3\vec{t}_3$, where $\vec{t}_1 = a(0,1/2,1/2)$, $\vec{t}_2 = a(1/2,0, 1/2)$ and $\vec{t}_3 = a(1/2,1/2,0)$ and  $n_1, n_2$ and  $n_3$ are integers.

By considering the cubic conventional cell with periodic boundary conditions, the group we use to to block-diagonalize the Hamiltonian is $O_h\times\{1,t_1,t_2,t_3\}$ (where the point group $O_h$ contains non-symmorphic elements), which has 192 group elements and 14 representations. 
Table \ref{table:nonlin} lists the dimension (degeneracy) of each IR, as well
as the size of the blocks of
the Hamiltonian and the number of pure spin ice states in each IR.

\begin{table}[ht]
\begin{tabular}{c|c|c|c } 
\hline 

IR & dimension & block size & number of pure
\\  
 & & & spin ice states \\
\hline 
A$_{1g}$ & 1& 383&4\\ 
A$_{2g}$ & 1&371&0 \\
A$_{1u}$ & 1&335&2 \\
A$_{2u}$ & 1&335&0 \\
E$_{g}$ & 2&774&4\\
E$_{u}$ & 2&682&2\\
T$_{1g}$ & 3&1081&2 \\
T$_{2g}$ & 3&1085&3 \\
T$_{1u}$ & 3&957&0 \\
T$_{2u}$ & 3&957&1 \\
X$_{1}$ & 6&2038&2 \\
X$_{2}$ & 6&2042&4 \\
X$_{3}$ & 6&2038&1 \\
X$_{4}$ & 6&2042&2 \\
\hline 
\end{tabular}
   \caption{Block-diagonalization of $H$ for 16 sites. The first column lists the IR's of the 
   symmetry group, the second column lists their dimension (degeneracy), the third column gives the size of each block ({\em i.e.} the number of basis states that belong to each dimension of each IR) and the final column lists the number of pure spin ice states belonging to each block.
    \label{table:nonlin}}
\end{table}

The unitary transformation that block-diagonalizes the Hamiltonian is found by generating the complete set of orthogonal basis states belonging to each IR. 90 of these states will be pure spin ice states (those states that satisfy the ice-rule on every tetrahedron), which appear in nearly all the representations of the group.  In addition to these pure spin ice states, there are many more states  where the ice rule is satisfied on some of the tetrahedra that also contribute to the two-in-two-out density.  These states occur in every block.

\section{Single Tetrahedron}
For comparison, we present results for a single tetrahedron. Since periodic boundary conditions are applied, the system contains two tetrahedra and so all interactions for a single tetrahedron occur twice (effectively doubling the eigenvalues). With a Hamiltonian of size $2^4$, this problem is easy to solve numerically, and block-diagonalization yields analytic results for some sectors. There are only 6 spin ice states altogether (all of them are pure spin ice states). 

\begin{table}[ht]
\begin{tabular}{c|c|c|c } 
\hline 
IR & dimension & block size & spin ice states
\\  
\hline 
A$_{1g}$ & 1& 1&1\\ 
A$_{2g}$ & 1&0&0 \\
A$_{1u}$ & 1&0&0 \\
A$_{2u}$ & 1&0&0 \\
E$_{g}$ & 2&3&1\\
E$_{u}$ & 2&0&0\\
T$_{1g}$ & 3&2&1 \\
T$_{2g}$ & 3&1&0 \\
T$_{1u}$ & 3&0&0 \\
T$_{2u}$ & 3&0&0 \\
\hline 
\end{tabular}
   \caption{Block-diagonalization of $H$ for 4 sites. The columns are the same as in Table \ref{table:nonlin}.
    \label{table:nonlin2}}
\end{table}

\begin{figure}[H]
\centering
\subfloat[${\cal J}_1 = -1$, ${\cal J}_4 = 0$]{
  \includegraphics[width=40mm]{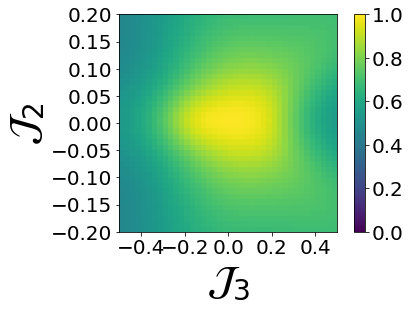}
}
\subfloat[${\cal J}_1 = -1$, ${\cal J}_3 = 0$]{
  \includegraphics[width=40mm]{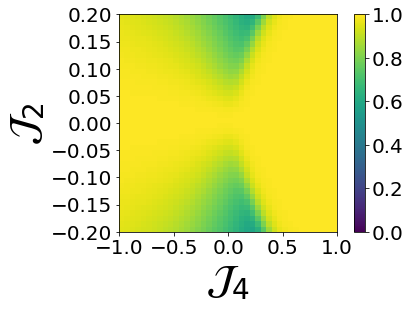}
}
\hspace{0mm}
\subfloat[${\cal J}_1 = -1$, ${\cal J}_2 = 0$]{
  \includegraphics[width=40mm]{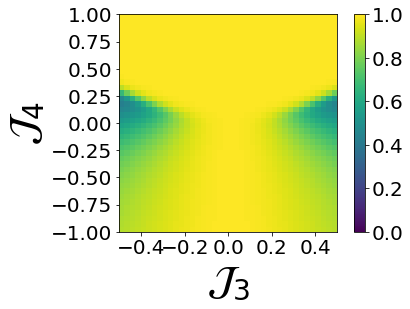}
}
\subfloat[${\cal J}_1 = -1$]{
  \includegraphics[width=40mm]{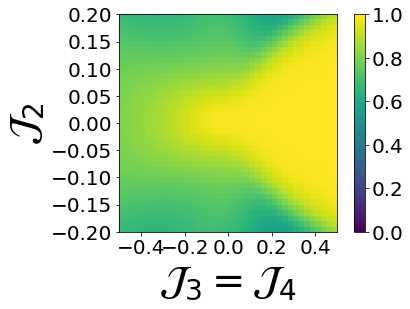}
}

\caption{The two-in-two-out density $n_{2I2O}$ for a single tetrahedron at $T=0.05$ for select
cuts within the parameter space of ${\cal J}_2$, ${\cal J}_3$ and ${\cal J}_4$.     \label{fig-app1}}
\end{figure}
\begin{figure}[H]
\centering
\subfloat[${\cal J}_1 = -1$, ${\cal J}_4 = 0$]{
  \includegraphics[width=40mm]{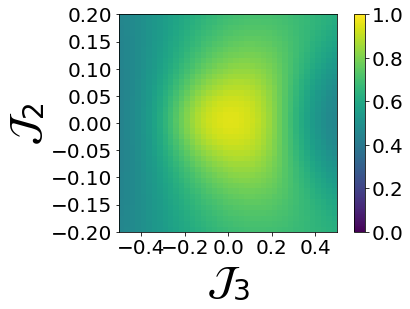}
}
\subfloat[${\cal J}_1 = -1$, ${\cal J}_3 = 0$]{
  \includegraphics[width=40mm]{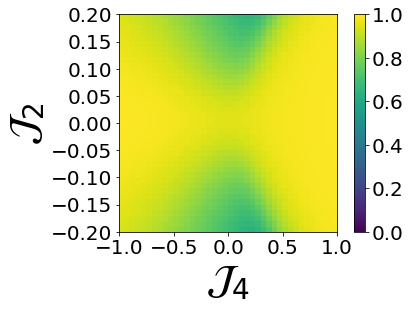}
}
\hspace{0mm}
\subfloat[${\cal J}_1 = -1$, ${\cal J}_2 = 0$]{
  \includegraphics[width=40mm]{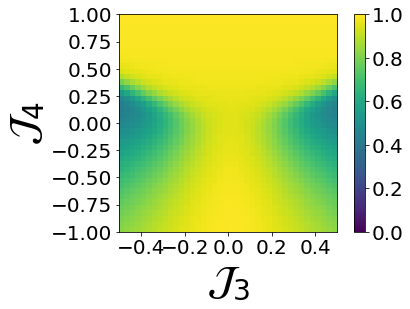}
}
\subfloat[${\cal J}_1 = -1$]{
  \includegraphics[width=40mm]{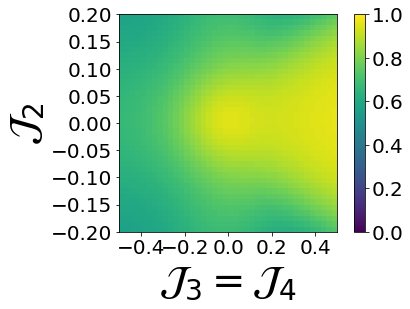}
}

\caption{The two-in-two-out density $n_{2I2O}$ for a single tetrahedron at $T=0.1$.   \label{fig-app2} }
\end{figure}
\begin{figure}[H]
\centering
\subfloat[${\cal J}_1 = -1$, ${\cal J}_4 = 0$]{
  \includegraphics[width=40mm]{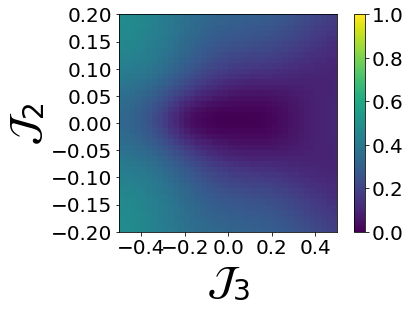}
}
\subfloat[${\cal J}_1 = -1$, ${\cal J}_3 = 0$]{
  \includegraphics[width=40mm]{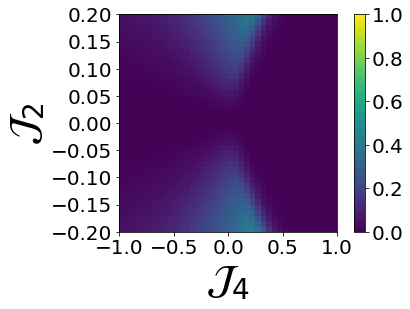}
}
\hspace{0mm}
\subfloat[${\cal J}_1 = -1$, ${\cal J}_2 = 0$]{
  \includegraphics[width=40mm]{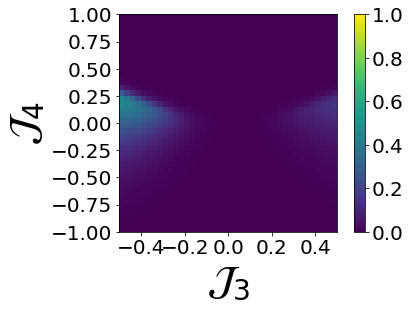}
}
\subfloat[${\cal J}_1 = -1$]{
  \includegraphics[width=40mm]{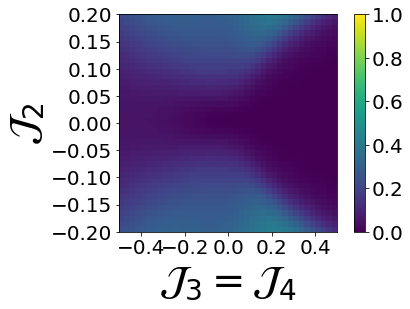}
}

\caption{The three-in-one-out/on-in-three-out density $n_{31/13}$ for a single tetrahedron at $T=0.05$.     \label{fig-app3}}
\end{figure}
\begin{figure}[H]
\centering
\subfloat[${\cal J}_1 = -1$, ${\cal J}_4 = 0$]{
  \includegraphics[width=40mm]{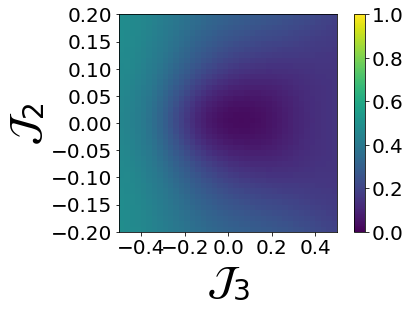}
}
\subfloat[${\cal J}_1 = -1$, ${\cal J}_3 = 0$]{
  \includegraphics[width=40mm]{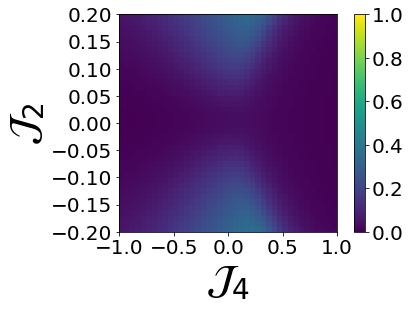}
}
\hspace{0mm}
\subfloat[${\cal J}_1 = -1$, ${\cal J}_2 = 0$]{
  \includegraphics[width=40mm]{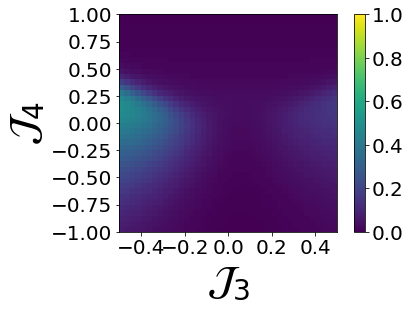}
}
\subfloat[${\cal J}_1 = -1$]{
  \includegraphics[width=40mm]{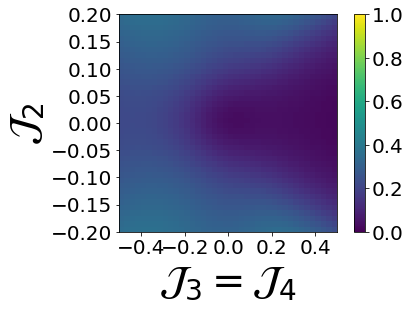}
}

\caption{The three-in-one-out/one-in-three-out density $n_{31/13}$ for a single tetrahedron at $T=0.1$ for select
cuts within the parameter space of ${\cal J}_2$, ${\cal J}_3$ and ${\cal J}_4$.     \label{fig-app4}}
\end{figure}
\begin{figure}[H]
\centering
\subfloat[${\cal J}_1 = -1$, ${\cal J}_4 = 0$]{
  \includegraphics[width=40mm]{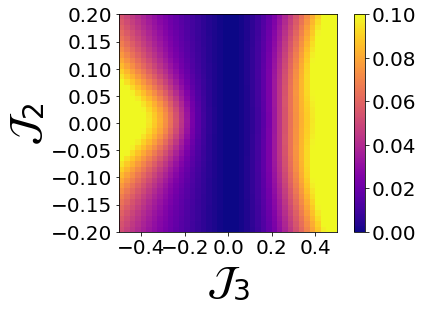}
}
\subfloat[${\cal J}_1 = -1$, ${\cal J}_3 = 0$]{
  \includegraphics[width=40mm]{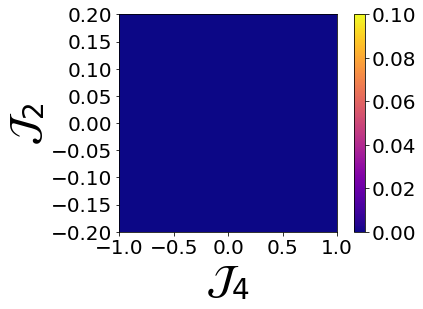}
}
\hspace{0mm}
\subfloat[${\cal J}_1 = -1$, ${\cal J}_2 = 0$]{
  \includegraphics[width=40mm]{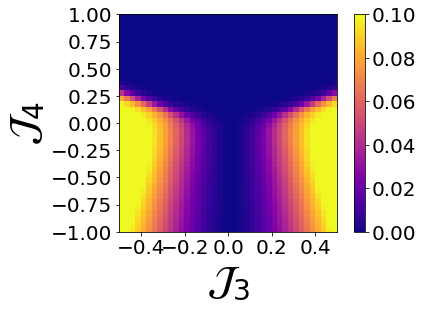}
}
\subfloat[${\cal J}_1 = -1$]{
  \includegraphics[width=40mm]{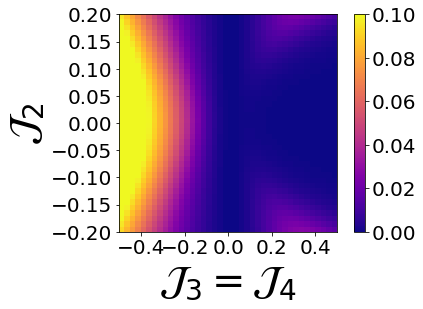}
}

\caption{The all-in-all-out density $n_{AIAO}$ for a single tetrahedron at $T=0.05$ .     \label{fig-app5}}
\end{figure}
\begin{figure}[H]
\centering
\subfloat[${\cal J}_1 = -1$, ${\cal J}_4 = 0$]{
  \includegraphics[width=40mm]{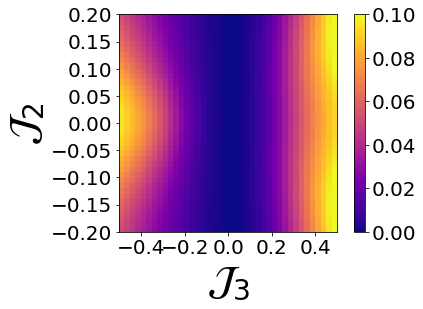}
}
\subfloat[${\cal J}_1 = -1$, ${\cal J}_3 = 0$]{
  \includegraphics[width=40mm]{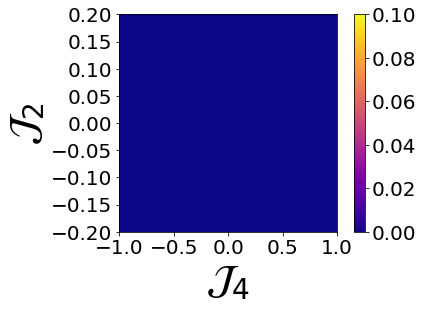}
}
\hspace{0mm}
\subfloat[${\cal J}_1 = -1$, ${\cal J}_2 = 0$]{
  \includegraphics[width=40mm]{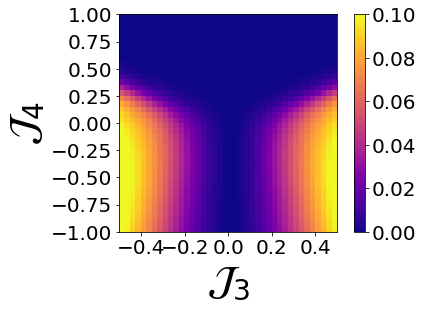}
}
\subfloat[${\cal J}_1 = -1$]{
  \includegraphics[width=40mm]{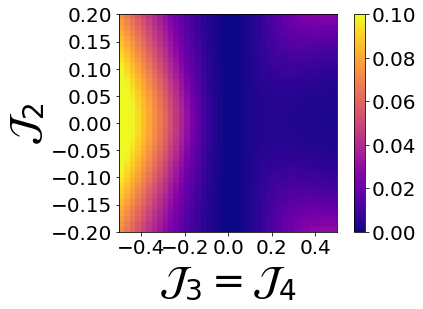}
}

\caption{The all-in-all-out density $n_{AIAO}$ for a single tetrahedron at $T=0.1$.     \label{fig-app6}}
\end{figure}

The spin ice density $n_{2I2O}$, the three-in-one-out/one-in-three-out density $n_{31/13}$ and the all-in-all-out density $n_{AIAO}$ (Eq.\ \ref{spin_ice}) for a single tetrahedron are
shown in Figs.~\ref{fig-app1} and \ref{fig-app2}, \ref{fig-app3} and \ref{fig-app4}, and \ref{fig-app5} and \ref{fig-app6}, respectively,
at $T=0.05$ and $T = 0.1$.

Symmetry and finite-size effects play a large role in the single-tetrahedron problem, which are especially evident when
${\cal J}_4 \neq 0$.  To analyze this problem, we consider the four terms in $H$ separately. As discussed in Section IIA, the operator $X_1$ has only diagonal matrix elements. Because
the operator $X_2$ contains single raising or 
lowering operators it {\em must} mix spin ice 
kets with other kets; likewise 
$X_3$ will have the same effect.  However, 
$X_4$ may connect spin ice states to either non-spin ice kets or to other spin ice kets.

As we have discussed, the general Hamiltonian can be block-diagonalized into blocks with basis kets belonging to the different IR's. It turns out that in the single-tetrahedron problem, after block-diagonalization, $X_4$ has only diagonal matrix elements, which means that $X_4$ will not mix the spin ice states appearing singly in each block with any other state.  Therefore,
large enough positive or negative values of ${\cal J}_4$ will produce a spin ice ground state
that will cause the value of $n_{2I2O}$ to be very close to one, hence there is no confinement of the pure spin ice phase for the parameter ${\cal J}_4$. This does not occur in the larger ($2^{16}$) system because $X_4$ has non-diagonal matrix elements in each block.
However the phase diagram in the space of parameters ${\cal J}_2$ and ${\cal J}_3$ is very similar to the larger system. 

\section{Exchange constants}
The relations between the exchange constants in Refs. \ \cite{Ross_physrevX_2011, petit_2016_PhysRevB.94.165153} and those used in (2) are:
\begin{eqnarray}
{\cal J}_1 & = & -3 J_{zz} \\
{\cal J}_2 & = & \frac{3}{\sqrt 2} J_{z \pm} \\
{\cal J}_3 & = & 3 J_{\pm\pm} \\
{\cal J}_4 &= & 6 J_{\pm}.
\end{eqnarray}

\bibliographystyle{apsrev4-1}
    \bibliography{citations}

\end{document}